\documentclass[a4paper,12pt,twoside]{article}
\usepackage[english]{babel}
\usepackage{rotating}
\usepackage{epsfig}
\usepackage[centertags]{amsmath}
\usepackage{array}
\usepackage{cite}
\usepackage{xspace}
\usepackage{amsfonts}
\usepackage{rotating}
\usepackage{float}
\def\3{\ss}
\newcommand{\PR}{Phys. Rev.\ }
\newcommand{\PRE}{Phys. Rep. \ }

\newcommand{\PL}{Phys. Lett.\ }
\newcommand{\NP}{Nucl. Phys.\ }
\newcommand{\ZP}{Z. Phys.\ }

\newcommand{\Coll}{Collaboration}
\newcommand{\CPC}{Comp. Phys. Comm. \ }
\newcommand{\xb}{X_{\mathrm{B}} }
\newcommand{\yb}{Y_{\mathrm{B}} }
\newcommand{\zb}{Z_{\mathrm{B}} }

\newcommand{\la}{\langle}
\newcommand{\ra}{\rangle}
\newcommand{\bec}{\begin{center}}
\newcommand{\eec}{\end{center}}
\newcommand{\cut}{\mathrm{cut}}

\def\as{\alpha_s}

\def\GeV2{\mathrm{GeV}^2}
\def\E{e^+e^-}
\def\ep{e^+p}

\begin{document}
\selectlanguage{english}


\title{
\begin{flushleft}
{\small\tt DESY 01-053\\ 
\vspace{-0.3cm} 
April 2001} 
\end{flushleft}  
\vspace{1cm}
\bf\LARGE Multiplicity moments in deep inelastic scattering at HERA\\}  

\author{ZEUS Collaboration}
\date{ }

\maketitle

\vfill
\centerline{\bf Abstract}
\vskip4.mm
\begin{center}
  \begin{minipage}{15.cm}
    \noindent
Multiplicity moments of  charged particles in deep inelastic
$\ep$ scattering  have been measured with the ZEUS detector at HERA using
an integrated luminosity of 38.4 pb$^{-1}$.
The moments for $Q^2>1000$ GeV$^2$ were studied in 
the current region of the Breit frame. 
The evolution of the moments was investigated   
as a function of restricted regions
in polar angle and, for the first time,
both in the transverse momentum and in absolute momentum
of final-state particles.
Analytic perturbative QCD predictions 
in conjunction with the  hypothesis of  
Local Parton-Hadron Duality (LPHD) 
reproduce the trends of the moments
in polar-angle regions, although some
discrepancies are observed. 
For the moments restricted either in transverse or absolute momentum,
the analytic results combined with the LPHD hypothesis show considerable
deviations from the measurements. The study indicates a large
influence of the hadronisation stage on the multiplicity distributions
in the restricted phase-space regions studied here, which is inconsistent
with the expectations of the LPHD hypothesis. 
\end{minipage}
\end{center}
\vfill
\thispagestyle{empty}

\newpage
%
%
%
%
                                                   %
\begin{center}                                                                                     
{                      \Large  The ZEUS Collaboration              }                               
\end{center}                                                                                       
  S.~Chekanov,                                                                                     
  M.~Derrick,                                                                                      
  D.~Krakauer,                                                                                     
  S.~Magill,                                                                                       
  B.~Musgrave,                                                                                     
  A.~Pellegrino,                                                                                   
  J.~Repond,                                                                                       
  R.~Stanek,                                                                                       
  R.~Yoshida\\                                                                                     
 {\it Argonne National Laboratory, Argonne, IL, USA}~$^{p}$                                        
\par \filbreak                                                                                     
  M.C.K.~Mattingly \\                                                                              
 {\it Andrews University, Berrien Springs, MI, USA}
% ...
\newpage                                                                                           
$^{\    1}$ now visiting scientist at DESY \\                                                      
$^{\    2}$ now at Univ. of Salerno and INFN Napoli, Italy \\                                      
% ...
                                                           %
                                                           %
\newpage   
                                                           %
                                                           %
\begin{tabular}[h]{rp{14cm}}                                                                       
$^{a}$ &  supported by the Natural Sciences and Engineering Research                               
          Council of Canada (NSERC)  \\                                                            
$^{b}$ &  supported by the FCAR of Qu\'ebec, Canada  \\                                            
$^{c}$ &  supported by the German Federal Ministry for Education and                               
          Science, Research and Technology (BMBF), under contract                                  
          numbers 057BN19P, 057FR19P, 057HH19P, 057HH29P, 057SI75I \\                              
$^{d}$ &  supported by the MINERVA Gesellschaft f\"ur Forschung GmbH, the                          
          Israel Science Foundation, the U.S.-Israel Binational Science                            
          Foundation, the Israel Ministry of Science and the Benozyio Center                       
          for High Energy Physics\\                                                                
$^{e}$ &  supported by the German-Israeli Foundation, the Israel Science                           
          Foundation, and by the Israel Ministry of Science \\                                     
$^{f}$ &  supported by the Italian National Institute for Nuclear Physics                          
          (INFN) \\                                                                                
$^{g}$ &  supported by the Japanese Ministry of Education, Science and                             
          Culture (the Monbusho) and its grants for Scientific Research \\                         
$^{h}$ &  supported by the Korean Ministry of Education and Korea Science                          
          and Engineering Foundation  \\                                                           
$^{i}$ &  supported by the Netherlands Foundation for Research on                                  
          Matter (FOM) \\                                                                          
$^{j}$ &  supported by the Polish State Committee for Scientific Research,                         
          grant no. 2P03B04616, 620/E-77/SPUB-M/DESY/P-03/DZ 247/2000 and                          
          112/E-356/SPUB-M/DESY/P-03/DZ 3001/2000\\                                                
$^{l}$ &  partially supported by the German Federal Ministry for                                   
          Education and Science, Research and Technology (BMBF)  \\                                
$^{m}$ &  supported by the Fund for Fundamental Research of Russian Ministry                       
          for Science and Edu\-cation and by the German Federal Ministry for                       
          Education and Science, Research and Technology (BMBF) \\                                 
$^{n}$ &  supported by the Spanish Ministry of Education                                           
          and Science through funds provided by CICYT \\                                           
$^{o}$ &  supported by the Particle Physics and                                                    
          Astronomy Research Council, UK \\                                                        
$^{p}$ &  supported by the US Department of Energy \\                                              
$^{q}$ &  supported by the US National Science Foundation                                          
\end{tabular}                                                                                      
                                                           %

\newpage
\pagenumbering{arabic}
\setcounter{page}{1}
\parindent3pt

\section{Introduction}
\label{sec:intr}

The hadronic final state in
deep inelastic  scattering (DIS) events is the result of a hard
partonic scattering initiated at large 
momentum transfers, $Q \gg \Lambda$, where $\Lambda$
is a characteristic QCD scale of the order of a few hundred MeV.   
The subsequent parton cascade  
is followed by a soft fragmentation process. The latter occurs 
with small momentum transfers which may be considered
to extend up to a value $Q_0$, which is   
a QCD cut-off above which perturbative methods can be applied.
There are two main approaches for describing the
hadronic final state. The first comprises analytic perturbative 
QCD calculations
based on the hypothesis of Local Parton-Hadron Duality (LPHD) \cite{AZIM}; 
this hypothesis states that parton-level QCD  
predictions are applicable for sufficiently 
inclusive hadronic observables,
without additional assumptions about hadronisation processes.
Therefore, the hadronic spectra are proportional to those of partons 
if the cut-off $Q_0$ is decreased towards a small value of $\Lambda$.  
The second  approach is formulated in Monte Carlo (MC) programs,   
which generate a  partonic 
final state according to the perturbative QCD picture and,  
below $Q_0\simeq 1$ GeV, hadronise  the partons 
on the basis of non-perturbative models.
The MC models are able to reproduce many detailed properties of the
hadronic  final state, but contain a large number of free parameters.
 
Multiplicity distributions and correlations between
final-state particles  
are an important testing ground for  
analytic perturbative QCD in conjunction with the LPHD hypothesis,  
as well as for MC  models  
describing the hadronic final state \cite{rew}.  
In this paper, a recent analysis of two-particle 
angular correlations
\cite{angul} is extended to a  study of  multiplicity distributions
measured in restricted phase-space regions of  neutral current DIS events.
The particle multiplicities  
are studied in terms of the 
normalised factorial moments\footnote{It may
be noted that, for $q=1$, $F_q=1$ by definition} 
$$
F_q(\Omega)=\la n(n-1)\ldots (n-q+1)\ra / \la n \ra^q,
\qquad q=2,3, \ldots ,  
$$
for a specified phase-space region of size $\Omega$. 
The number, $n$, of particles is measured inside  $\Omega$
and angled brackets $\la\ldots\ra$ denote 
the average over all events. The factorial
moments, along with cumulants \cite{cum} and bunching parameters \cite{bp},
are convenient tools  to characterise the multiplicity
distributions when $\Omega$ becomes small.
For uncorrelated particle production within $\Omega$, 
Poisson  statistics
holds  and $F_q=1$ for all $q$. Correlations
between particles lead to a broadening of the multiplicity distribution 
and to dynamical fluctuations. In this case, the normalised factorial 
moments increase  
with decreasing $\Omega$. If  the  rise follows  a power law,  
this  effect is frequently called ``intermittency'' \cite{bias}.  

The LPHD hypothesis, originally suggested for single-particle spectra, 
presently has experimental support in $\E$ annihilations \cite{QCDh}. 
However, the success of LPHD is less  evident for
the moments of inclusive single-particle densities 
at HERA energies \cite{mult}
and  two-particle angular correlations \cite{angul}.  
In contrast to the single-particle densities and 
global event-shape variables, 
the factorial moments are most directly affected by
inclusive {\em many}-particle densities.
This can be seen from  the relation 
$\la n(n-1)\ldots (n-q+1)\ra =
\int_{\Omega}  \rho^{(q)}(p_1,\ldots p_q) dp_1\ldots dp_q$ between 
the factorial moments in a region $\Omega$ and the inclusive
$q$-particle density, 
$\rho^{(q)}(p_1,\ldots p_q)=(1/\sigma_{tot})\> d^{q}\sigma /dp_1\ldots dp_q$.
The latter represents  the  
probability to detect $q$ particles with  momenta
$p_1,\ldots p_q$, irrespective of the presence of any other particles.   
Thus the factorial moments are an important tool to
study such  densities, for which the applicability of    
LPHD remains questionable even for  higher $\E$ energies at 
LEP \cite{fluc_l3, fluc_del}. 
Deviations of the measured moments from
theoretical predictions  would imply
the necessity for further refinements of perturbative QCD 
calculations and/or the failure of the 
LPHD hypothesis. In the latter case, the  
magnitude of such deviations 
can  shed light on details
of the hadronisation processes.    
 
\section{Analytic QCD results}
\label{sec:anal}

The dominant source of
particle production inside a jet 
is gluon splitting in the QCD cascade, 
where the presence of  a gluon enhances the probability for
emission of another gluon nearby in momentum space.
This leads  to inter-parton  
correlations and non-Poissonian statistics for the 
multiplicity distributions in restricted  phase-space intervals  
where partons are counted. 
Thus the factorial moments, $F_q$, are expected to 
deviate from unity. 

There exist a number of analytic QCD predictions for the moments
of the parton multiplicity distributions obtained in 
the Double Leading Log Approximation (DLLA), which are  
discussed below.    

\subsection{Multiplicity-cut moments}

\begin{figure}
\begin{center}

\vspace{2.5cm}

\mbox{\epsfig{file=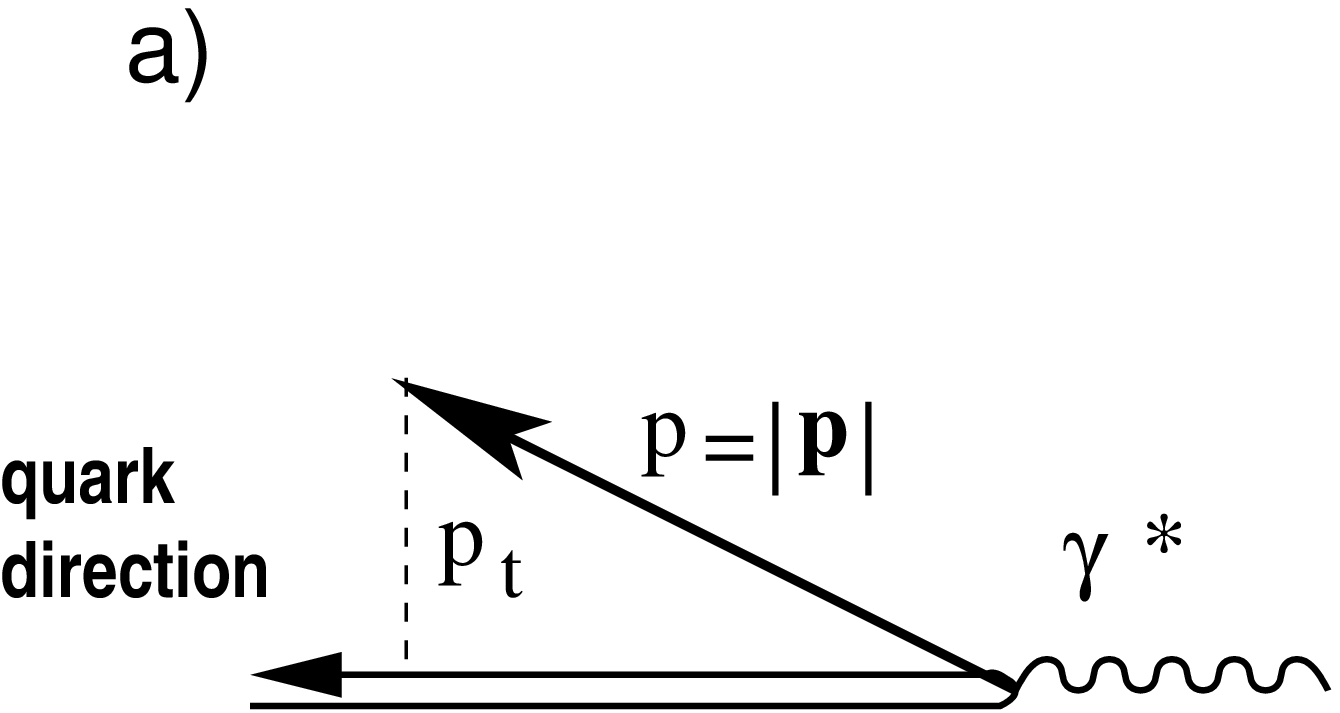, width=7.5cm}}

\vspace{3.0cm}

\mbox{\epsfig{file=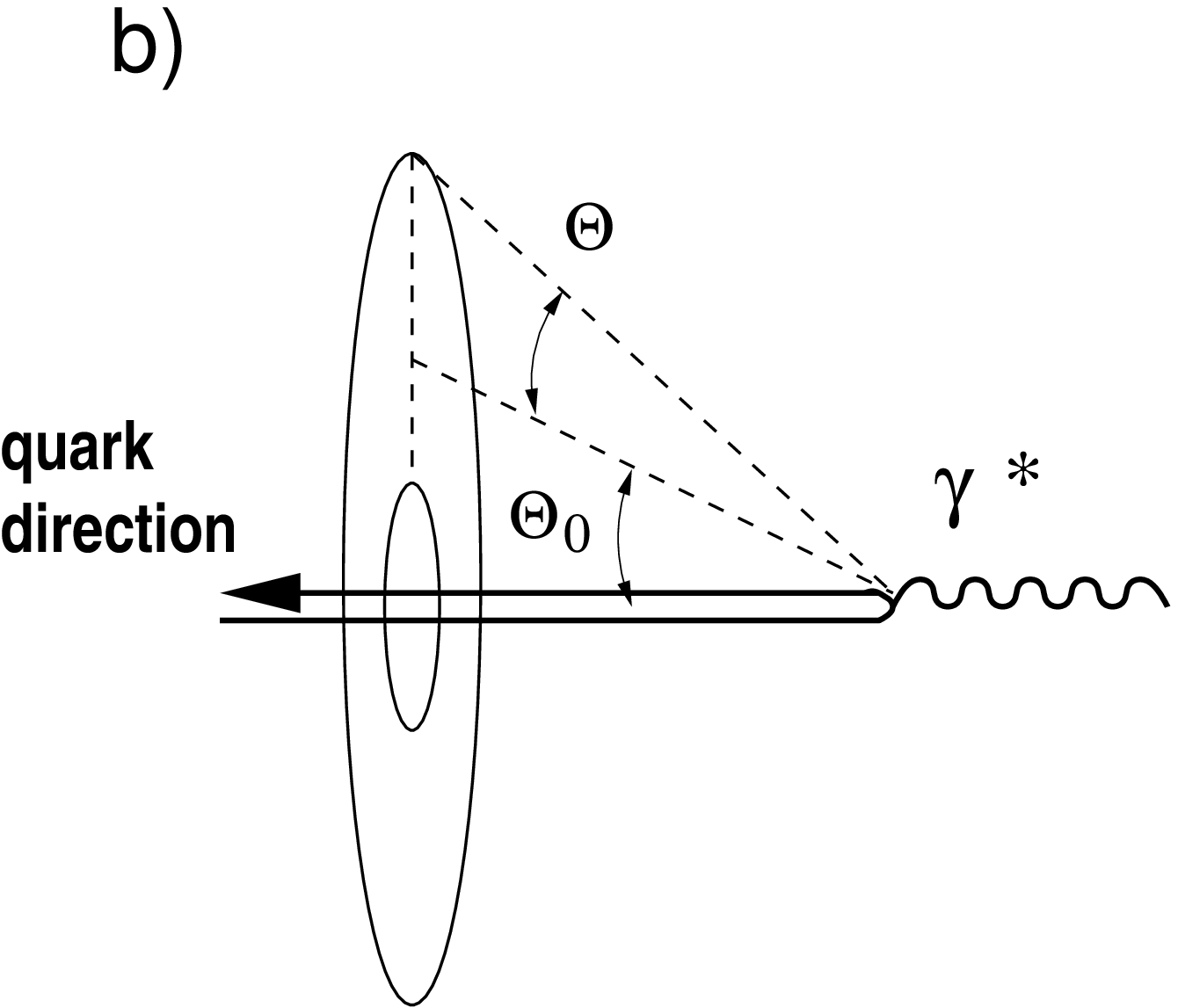, width=8.0cm}}

\caption{A schematic representation of the measurements of the
factorial moments in DIS: (a) restricted in
the transverse momentum, $p_t$,  defined   
with respect to the outgoing-quark direction  
and absolute  momentum, $\mid {\bf  p } \mid $; (b) in
polar-angle rings of width 2$\Theta$.} 
\label{fig:0}
\end{center}
\end{figure}

The normalised factorial moments
of the multiplicity distributions of gluons   
which are restricted in either 
transverse momentum $p_{t}<p_{t}^{\cut}$ 
or absolute  momentum
$p\equiv \mid {\bf p } \mid <p^{\cut}$ are expected, respectively,  to  
have the following behaviour \cite{ochs_pt}:
\begin{eqnarray}
F_q(p_t^{\cut}) & \simeq &
1 + \frac{q(q-1)}{6}\>  \frac{\ln(p_t^{\cut}/Q_0)}{\ln(E/Q_0)}, 
\nonumber \\[-0.1cm]    
\\[-0.1cm] 
F_q(p^{\cut}) & \simeq & C(q) > 1, \nonumber
\label{eq01}
\end{eqnarray}
where $E$ is the initial 
energy of the outgoing quark that radiates the gluons and
$C(q)$ are constants depending on $q$.   
The maximum transverse momentum, $p_{t}^{\cut}$, and 
the $p_t$ values are defined with respect to the
direction of this quark 
(see Fig.~\ref{fig:0}(a) for a schematic representation 
in the case of DIS).
 
Equations (1) indicate  that
there  are positive correlations between partons, because the
factorial moments are larger than unity. For small $p_t^{\cut}$ values,
however, the correlations vanish due to the presence of   
an angular ordering of the partons in the jet.   
The existence of the angular ordering
leads  to the absence of branching processes with secondary gluon
emission at small $p_{t}^{\cut}$. This
ultimately leads to the suppression of the correlations  in this
phase-space region and to independent parton emission\footnote{ 
Note that this effect is similar to
multiple photon bremsstrahlung in QED.} 
($F_q\simeq 1$ for $p_t^{\cut} = Q_0$).
On the other hand, the distribution
of soft gluons with limited absolute momenta, $p<p^{\cut}$, remains 
non-Poissonian  for any small value of $p^{\cut}$.

Note that the theoretical formulae (1) can be regarded as asymptotic
calculations, i.e. they are valid at small $p_t^{\cut}$ and $p^{\cut}$.  
Therefore, the calculations  should be  considered only as 
qualitative predictions 
when compared to the data using the LPHD hypothesis.     

\subsection{Angular moments}

Gluon  multiplicities in polar-angle rings 
around the outgoing-quark direction 
are expected to exhibit a distribution  which is  substantially wider than
for Poisson statistics \cite{drem,brax,ochs}. 
The moments rise with increasing  values of the phase-space 
variable  $z$ as  
\begin{equation}
\ln \frac{F_q(z)}{F_q(0)}=z\> a\> (1-D_q)(q-1),
\label{ang}
\end{equation}
where $z$ and the  constant $a$
are defined as:  
\begin{eqnarray}
z & = & a^{-1} \ln \frac{\Theta_0}{\Theta},
\nonumber \\[-0.1cm] 
\\[-0.1cm] 
a & = & \ln \frac{E\Theta_0}{\Lambda}. \nonumber 
\label{angdef}
\end{eqnarray}
The angle $\Theta_0$ is the half 
opening-angle of a cone  around the outgoing quark radiating the gluons,
$E$ is its energy, and $\Lambda$ is an
effective QCD scale ($\Lambda\neq\Lambda_{\overline{MS}}$).
A schematic representation  of the 
variables is   shown in Fig.~\ref{fig:0}(b), in which  
a window in the angular half-width $\Theta$ 
is subtended by the ring
centered at $\Theta_0$.  
The factors $D_q$ are the so-called R\'{e}nyi 
dimensions \cite{renyi} known from
non-linear physics.     
A decrease of the angular window $\Theta$ corresponds
to an increase of the variable $z$. 
The maximum possible  
phase-space region ($\Theta = \Theta_0$), corresponding  to $z=0$,
is used to calculate the factorial moments, $F_q(0)$, which are necessary 
for normalisation in Eq.~(2). Such normalisation removes the  
theoretical uncertainty in the absolute values of $F_q(z)$
and thus allows a comparison of the predicted $z$-dependence
of the moments with experimental results.

For independent particle production in a restricted range of $2\Theta$, 
$D_q=1$ and $F_q(z)=F_q(0)$ in Eq.~(2). 
In contrast, the analytic QCD expectations for $D_q$ obtained in the DLLA  
are as follows \cite{drem,brax,ochs}:

\begin{itemize}
\item 
if the strong coupling constant, $\as$, 
is fixed at scale $Q\simeq  E\Theta_0$,
then for  large angular bins (i.e. small  $z$),  
\begin{equation}  
D_q = \gamma_0\frac{q+1}{q} , 
\label{an1} 
\end{equation}
where $\gamma_0=\sqrt{2\,C_{\mathrm{A}}\as/\pi}$ is
the anomalous QCD dimension,  
and $C_{\mathrm{A}}=3$   
is the gluon colour factor. 
The $\as$ value is calculated at the one-loop level with 
$n_f=3$ (number of flavours).     
Such a behaviour of the moments corresponds  to the expectations from 
intermittency, which can be
interpreted in terms of the fractal  
structure of the gluon splittings \cite{rew}.         
\end{itemize}

For running-$\as$ values, 
the $D_q$ can be approximated
by different  expressions, depending on the treatment  
of the non-leading DLLA contributions to the moments:  

\begin{itemize}
\item 
for Dokshitzer and Dremin \cite{drem}:  
\begin{equation}    
D_q\simeq\gamma_0\,\frac{q+1}{q}\left(1 + \frac{q^2+1}{4q^2}z\right);  
\label{an2}
\end{equation}

\item 
for Brax et al. \cite{brax}:
\begin{equation}    
D_q\simeq2\,\gamma_0\,\frac{q+1}{q}\left(\frac{1-\sqrt{1-z}}{z}\right);  
\label{an3}
\end{equation}

\item
for Ochs and Wosiek \cite{ochs}:  
\begin{equation}    
D_q\simeq2\,\gamma_0\,\frac{q-w(q,z)}{z(q-1)}, \qquad
      w(q,z)=q\sqrt{1-z}\left(1-\frac{\ln(1-z)}{2q^2}\right);  
\label{an4}
\end{equation}

\item 
an  estimate for $D_q$ 
which includes a correction from the Modified Leading Log Approximation 
(MLLA) has also been obtained by Dokshitzer and Dremin \cite{drem}.
In this case, Eq.~(5) remains valid, but $\gamma_0$ 
is replaced
by an effective $\gamma_0^{\mathrm{eff}}(q)$ which depends on $q$. 
\end{itemize}

All the theoretical calculations quoted above are performed
for partons. Therefore, for comparisons  of the predictions  with
data, LPHD is assumed. 

\section{Experimental set-up}
\label{sec:es}
 
The data were  taken in 1996 and 1997
using the ZEUS detector at HERA. 
During this period, the energy of the positron
beam, $E_e$, was $27.5$ GeV  and that of the proton beam was  $820$ GeV.
The integrated luminosity used for the 
present study was $38.4\pm 0.6$ pb$^{-1}$.
 
ZEUS is a multi-purpose  
detector described in detail elsewhere \cite{zeus_det}. 
Of particular importance for  the
present study  are the central tracking detector (CTD),
positioned  in a  1.43~T solenoidal magnetic field,     
and the calorimeter.

The CTD is  a cylindrical
drift chamber with nine superlayers covering the polar
angle\footnote{
The ZEUS coordinate system is a right-handed Cartesian system, with
the $Z$ axis pointing in the proton beam direction, and the $X$ axis
pointing left towards the centre of HERA. The coordinate origin is at the
nominal interaction point.
The polar angle $\theta$ is defined with
respect to the positive $Z$-direction.
The  pseudorapidity, $\eta$,
is given by $-\ln ( \tan \frac{\theta}{2} ) $.}
region $15^o < \theta \leq 164^o$ and the
radial range 18.2-79.4 cm. Each superlayer consists
of eight anodes. The transverse-momentum resolution
for charged tracks traversing all CTD layers is 
$\sigma(p_{\bot})/p_{\bot} = 0.0058 p_{\bot}  
\oplus 0.0065 \oplus  0.0014/p_{\bot}$,  
with $p_{\bot}$ being the track-transverse momentum (in GeV). 
The single hit efficiency of the CTD is greater than $95\%$. 

The CTD 
is  surrounded by the uranium-scintillator calorimeter (CAL), which    
is longitudinally segmented into electromagnetic 
and hadronic sections. The relative energy resolution of the calorimeter 
under test beam conditions 
is $0.18/\sqrt{E}$ for electrons and 
$0.35/\sqrt{E}$ for hadrons (with $E$ in GeV).
The CAL cells provide time measurements with a time resolution 
below 1 ns for energy deposits larger than 4.5 GeV.

\section{Data selection}
\label{sub:sel}

\subsection{Event kinematics}

The basic event kinematics  of  DIS  processes are
determined by the negative squared four-momentum transfer
at the positron vertex,
$Q^2=-q^2=-(k-k^{'})^2$ ($k$ and $k^{'}$ denote the four-momenta of the
initial- and final-state positrons, respectively)
and the Bjorken
scaling variable $x=Q^2/(2 P\cdot  q)$,
where $P$ is the
four-momentum of the proton.
The fraction of the energy transfered 
to the proton in its rest frame, $y$,
is related to these two variables
by  $y\simeq Q^2/s x$,
where $\sqrt{s}$ is the positron-proton
centre-of-mass energy.

The moments were studied in the Breit
frame \cite{Br},  which provides a natural system to separate
the radiation from  the outgoing
struck quark from  the proton remnants.
In this frame,
the exchanged virtual boson with virtuality $Q^2$ is completely
space-like and has a momentum $q=(q_0,q_{\xb},q_{\yb},q_{\zb} )=(0,0,0,-Q)$.
In the quark-parton model, the
incident quark has $p_{\zb}=Q/2$
and the outgoing
struck quark carries $p_{\zb}=-Q/2$.  
All particles with negative $p_{\zb}$
form the current region. These  particles
are produced by the fragmentation   of
the struck quark, so that the current region is analogous to a
single hemisphere in $\E$ annihilations.
The variables $p_t=\sqrt{p_{\xb}^2 + p_{\yb}^2}$
and $\Theta_0$ were determined with respect to
the negative $\zb$-direction,
which, for high $Q^2$ events, is a good estimate of the 
outgoing struck-quark direction.
The moments in the current region of the Breit frame 
were  measured in the kinematic region 
$Q^2>1000$ GeV$^2$, which is sufficiently
high to avoid  major contributions  from the boson-gluon 
fusion (BGF) processes  
(see Section~\ref{sec:ss}).

\subsection{Event and track selection} 

The kinematic variables  $x$ and $Q^2$ were  
determined by a combination of methods:  
(i) the measurement of the energy and angle of the scattered 
positron (denoted by the subscript $e$);   
(ii) the double angle (DA) method \cite{dam} and (iii) 
the Jacquet-Blondel (JB)  method \cite{damj}, depending on which is
most appropriate for a particular cut.   
In the DA method, the 
variables $x$, $Q^2$ and $y$ were reconstructed 
using the angles of the scattered  positron  and the hadronic energy flow.
In the JB method, used only in the event selection,
the kinematic variables  were  determined entirely from the
hadronic system.  

The boost vector from the  
laboratory to the Breit frame was  determined using the DA method,
which is less sensitive to systematic uncertainties in the 
energy measurement than the other methods. 
In the reconstruction of the Breit frame, 
all charged particles were assumed to have the pion mass.  

The triggering and online event selections were  
identical to those used in a previous publication \cite{trigg}. 
To preselect  neutral current DIS events,   
the following cuts were applied:

\begin{enumerate}
 
\item[$\bullet$]
$E_e^{'}\geq 10$ GeV, $E_e^{'}$ being the energy of the scattered positron; 
 
\item[$\bullet$]
$Q^2_{\mathrm{DA}}\geq 1000$ GeV$^2$; 

\item[$\bullet$]
$35\> \leq \> \delta=\sum E_i(1-\cos\theta_i) \> \leq \> 60$ GeV, where
$E_i$ is the energy of the $i^{\mathrm{th}}$  calorimeter cell and 
$\theta_i$ is its polar angle with respect to the beam axis; 
 
\item[$\bullet$]
$y_{e} \> \leq \> 0.95$; 
 
\item[$\bullet$]
$y_{\mathrm{JB}} \> \geq \> 0.04$;

\item[$\bullet$]
$Z$ coordinate  of the  reconstructed  vertex position, as
determined from the
tracks fitted to the vertex, in the range
$-40 < Z_{\mathrm{vertex}}  < 50$ cm;
 
\item[$\bullet$]
a timing cut requiring that the event time as measured by the CAL  
be consistent with an $\ep$  interaction.   

\end{enumerate}

The resulting data set contained 7369 events.

\vspace{0.4cm}
 
The present study is based on the tracks reconstructed with the CTD 
and fitted to the event vertex.
The scattered positron  was removed from the track sample.
In addition, the following cuts to select tracks 
were imposed:    
 
\begin{enumerate}
 
\item[$\bullet$]
$p_{t}^{lab}>0.15$  GeV,  where $p_{t}^{lab}$ is the 
transverse momentum of the track;  
 
\item[$\bullet$]
$\mid \eta \mid < 1.75$.
\end{enumerate}

These cuts restrict the data 
to  a well understood, high-acceptance region of the CTD.

\section{Event simulation}
\label{sec:esim}
 
Neutral current DIS events were   
generated with ARIADNE 4.10 \cite{ARD}
with the high $Q^{2}$ modification developed during 
the HERA Monte Carlo workshop \cite{ARDMOD} 
using tuned  parameters \cite{bruk}.
This  MC program is based on the colour-dipole model,
in which gluons are emitted from the colour field between quark-antiquark
pairs, supplemented  with the BGF process. 
The hadronisation was  simulated using the Lund string model
as implemented in JETSET \cite{je}. 
Charged hadrons with a mean lifetime longer than $10^{-8}$ seconds
were considered as stable. The CTEQ4D \cite{CTEQ4D} 
parameterisation of the proton parton distribution functions  was  used.
The events were generated without any cuts on kinematic variables
and track quantities.   
The MC events  obtained  by   this procedure 
defined  the generator-level sample. 

To obtain a 
detector-level sample,
events were  generated with  the
DJANGO 6.24 \cite{django} program based on HERACLES 4.5.2 \cite{heracles}, 
in order to  incorporate first-order electroweak corrections.
The parton cascade and hadronisation were then simulated with ARIADNE. 
The events were  then  processed  through a simulation of the detector
using GEANT 3.13 \cite{geant} to take into account    
particle interactions with the detector material, particle decays,
event and track migrations,
resolution, acceptance of the detector and  event
selection.
The detector-level MC events were processed with the same
reconstruction program as was used  for the data.

In addition to ARIADNE,  the LEPTO 6.5 \cite{LEP}
and HERWIG 5.9 \cite{HEW} generators with tuned parameters \cite{bruk} 
were  also used to compare with the data. 
The  parton cascade in LEPTO is based on a matrix-element calculation
matched to parton showers according to the DGLAP equations. 
As in ARIADNE, the hadronisation in LEPTO is simulated with JETSET. 
HERWIG  also has  a  parton shower based on DGLAP evolution, but 
parton emissions are  ordered in angle, in contrast to
LEPTO, which orders the parton emissions in invariant mass with an additional
angular constraint to ensure coherence. The hadronisation in HERWIG is     
described by the cluster fragmentation model \cite{cluster}.   

The Bose-Einstein (BE) interference between
identical particles can produce additional correlations
which distort the behaviour of  
the moments measured in small phase-space regions.
By default, the BE interference is turned off in JETSET;  it is absent  
in HERWIG. The effect of the BE correlations on the present 
results is considered in the next section.  

\section{Correction procedure }
\label{sec:cp}

The bin-by-bin correction factors,
$\mathcal{C}_q =
\mathcal{A}_q^{\mathrm{gen}} / \mathcal{A}_q^{\mathrm{det}}$, 
for  a given kinematic region
in $Q^2$ were evaluated using the ARIADNE 4.10 MC 
event samples, separately for each
observable, 
$\mathcal{A}_q=F_{q}(p_t^{\cut}),F_{q}(p^{\cut})$ and $F_q(z)/F_q(0)$,  
where $\mathcal{A}_q^{\mathrm{gen}}$ ($\mathcal{A}_q^{\mathrm{det}}$)  
was calculated at  the generator (detector) level. 
The factors $\mathcal{C}_q$ correct the data for the detector effects
and event-selection cuts as well as for the effects that 
are not included in the analytic calculations, namely 
(i) Dalitz decays of the $\pi^0$; (ii) leptons in the final state and   
(iii) initial-state photon  radiation. 

The correction
factors  for the $p_t^{\cut}$ and $p^{\cut}$ moments
are approximately flat and vary between about 1.0
and 1.3, except for the smallest momentum cuts 
and the highest-order $p_t^{\cut}$ and $p^{\cut}$ moments, where 
the correction factors 
are as large as  $1.7$.
The correction factors for the $F_q(z)/F_q(0)$ moments
vary smoothly from  1.0 to  1.5. The contribution of  
Dalitz pairs to the correction factors  is negligible,
except for the smallest phase-space regions in 
$p_t^{\cut}$, $p^{\cut}$ and $\Theta$, where it does not exceed
$20\%$. 

The analytic QCD calculations do not include BE interference.   
To take into account this phenomenon,  
the correction factor, BE$_q$,  was  calculated for each 
data point 
\begin{equation}
\mathrm{BE}_q =
\frac{\mathcal{A}_q^{\mathrm{gen}}}
{\mathcal{A}_q^{\mathrm{BE}, \> \mathrm{gen}}} \> \> ,
\label{be}
\end{equation}
where $\mathcal{A}_q^{\mathrm{gen}}$ 
($\mathcal{A}_q^{\mathrm{BE}, \> \mathrm{gen}}$) 
are the moments at  the generator level
without (with) the BE effect.  
For this correction, the BE
effect according to a Gaussian parameterisation with JETSET
default parameters was used. As a cross check,   
the BE correction was  determined  from the ZEUS data and also using  
the parameters obtained by the H1
Collaboration \cite{H1be}.  
The differences between the correction factors BE$_q$ for these
three sets of parameters were small. The values of the BE correction factors
are discussed in Section~\ref{sec:results}.   

\section{Systematic uncertainties}
\label{sec:ss}

The overall systematic uncertainty for each moment
was determined from the following 
uncertainties added in quadrature: 

\begin{enumerate} 

\item[$\bullet$]
the choice of the MC models. HERWIG was used 
in place of ARIADNE to determine the correction factors.
This uncertainty  was  typically $45\%$
of the total systematic uncertainty;

\item[$\bullet$]
event reconstruction and selection. Systematic
checks were  performed by varying the cuts on 
$\delta$,  $y_{e}$, $y_{\mathrm{JB}}$ 
and the vertex position requirement 
($40 \> \leq \> \delta \> \leq \>  55$ GeV, 
$y_{e} \> \leq \>  0.85$, $y_{\mathrm{JB}} \> \geq \>  0.05$
and $-35 < Z_{\mathrm{vertex}}  < 45$ cm). The overall uncertainties
associated with the event selection were typically $25\%$
of the total systematic uncertainty; 
 
\item[$\bullet$] 
track reconstruction and selection.   
Tracks were required to have transverse momenta larger than $0.2$ GeV 
and $\mid \eta \mid < 1.5$.  
Tracks that reach at least the 
third CTD  superlayer were  used.  
These  uncertainties  were  typically $30\%$
of the total systematic uncertainty. 

\end{enumerate}

As an example, the  total systematic uncertainty of $F_2(p_t^{\cut})$
for the smallest $p_t^{\cut}$ value used in this analysis 
is approximately  $6\%$.  
For the large $p_t^{\cut}$ regions, a typical value of the systematic 
uncertainty is $1\%$. 

To ensure that the two particles in a pair were  resolved,  the
angle  between two tracks  in the current region of the Breit frame
should not be smaller than two  degrees \cite{angul}.
This limit  determines  the minimum size of    
phase-space regions used in  this analysis.

The kinematic variables $p_t$, $\Theta$ and $\Theta_0$
have to be measured with respect to the outgoing-quark direction,
which coincides  with the $\zb$-direction of the Breit frame only  
for the zeroth-order $\gamma^*q$ scattering. 
The first-order QCD effects, 
BGF ($\gamma^*g\to q\overline{q}$) and
QCD Compton ($\gamma^*q \to qg$),
can lead to two hard partons that  are not collinear
to the $\zb$-direction.
For these events, the determination of the kinematic
variables with respect to the $\zb$ axis of the Breit frame
is inappropriate, especially when these two hard  partons
escape the current region.   
Using the LEPTO MC model, it has been  determined
that, at $Q^2>1000$ GeV$^2$, the number of  BGF and QCD Compton events 
without hard partons in the current
region, relative to the
total number of events, is  
$\leq$ 15\%.    
As a systematic check, the factorial
moments  have been calculated in kinematic variables
with respect to the thrust axis
determined from the charged particles in the current region of the
Breit frame. The results agree  
with those for which the negative $\zb$-direction of the
Breit frame is used.

\section{Results}
\label{sec:results}

The charged-particle multiplicities  measured in the entire  current
region without additional momentum cuts  on the charged particles 
were  compared to the previous ZEUS measurements \cite{mult} 
and to $\E$ data \cite{tasso}.
For the present measurement,
the corrected average value of $Q^2$
is  $\la Q^2\ra \simeq 2070$ GeV$^2$,
which corresponds to an  
energy $\sqrt{ \la Q^2 \ra } \simeq  46$ GeV.
The corrected
average charged-particle multiplicity in the current region,
$5.76\pm 0.29$, where the error represents statistical and systematic
uncertainties added in quadrature,  
agrees with the previous  ZEUS
measurement, $6.01\pm 0.46$,  
for a  similar value of $\la Q^2 \ra$.
This average multiplicity
is lower  than that found in  a  single hemisphere
of  $\E$ at $\sqrt{s}=44$ GeV, which is $7.58\pm 0.37$ \cite{tasso}.
The second-order factorial moment $F_2$
in a single hemisphere of
$\E$ annihilations is\footnote{
The second-order moment in a single hemisphere of
$\E$ annihilation can be estimated
from the dispersion $D=\sqrt{\la n^2\ra - \la n\ra^2}$
and the average multiplicity $\la n\ra$ \cite{tasso}
using the relation $F_2= 1+ D^2 \> \la n \ra^{-2} - \la n \ra^{-1}$.}
$F_2=1.036\pm 0.026$,
which is consistent with the value $1.051\pm 0.018$  obtained
in this study of the entire current  region.   
This difference in the average multiplicity  can be attributed to
a  migration of particles between the
current and target regions  due to BGF and QCD Compton processes,
an effect which has been theoretically quantified
\cite{fbche} and directly measured by ZEUS \cite{angul}. Such a migration
leads to current-target anti-correlations which do not vanish
even at the high $Q^2$ values studied  here. This is unlike the
$\E$ annihilation processes, where  weak and positive
forward-backward correlations have been observed
at a similar energy \cite{tasso}.

\subsection{Multiplicity-cut moments}

\begin{figure}
\begin{center}
\vspace{1.0cm}
\mbox{\epsfig{file=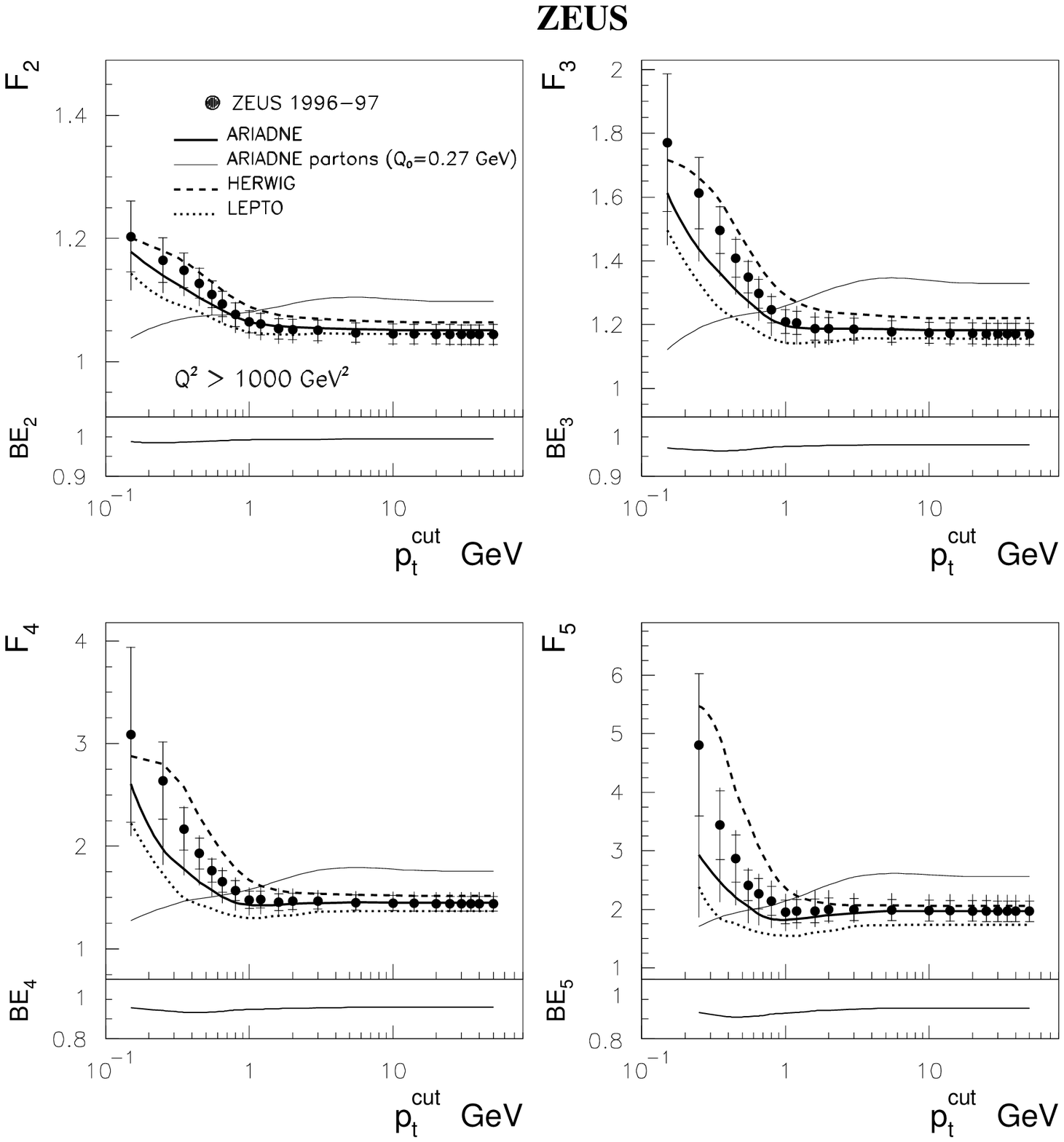, height=20.0cm}}
\vspace{-3.0cm}
\caption{
Factorial moments for charged particles  
in the current region of the Breit frame  as a function of $p_{t}^{\cut}$, 
compared to  Monte Carlo models. The thin lines show the  parton-level
prediction of ARIADNE with $Q_0=0.27$ GeV 
and the thick solid lines are the hadron-level 
predictions of the same model.   
The dashed and dotted lines show the hadron-level predictions
of HERWIG and LEPTO, respectively.
The inner error bars
are statistical uncertainties; the outer are statistical and systematic
uncertainties  added in quadrature. The correction factors for the BE 
effect are shown below each plot.
}  
\label{fig:6}
\end{center}
\end{figure}

The multiplicity moments of order $q=2,\ldots ,5$ as a function of
$p_t^{\cut}$ are shown in  
Fig.~\ref{fig:6}.
The $p_t^{\cut}$ moments were  calculated with
respect to the negative $\zb$-direction of the Breit frame.
As $p_t^{\cut}$ decreases  below 1 GeV,
the moments rise. This disagrees 
with the DLLA theoretical calculations, which predict  
that the moments approach unity (see Eq.~(1)).   
On the other hand, Monte Carlo models show the same
trend as the data, although some differences in slopes are apparent. 
LEPTO exhibits  smaller  values of the moments than for the data 
at small  $p_t^{\cut}$ values, while 
HERWIG overestimates the moments.
ARIADNE gives the 
best overall description of the correlations.

Figure \ref{fig:6} also shows the factors BE$_q$ from  
Eq.~(8) used  to correct the data for the BE interference. 
The largest correction was found in the
case of moments of high orders ($q=4,5$)
for  small $p_t^{\cut}$ cuts  ($<0.5$ GeV). 

To investigate contributions of the hadronisation processes  
to the moments,
further samples of the ARIADNE MC were generated: (1) considering
neutral hadrons in addition to charged particles; (2) using  only
hadrons coming from the Lund string fragmentation without
resonance decays. For these two cases, 
the moments have an even steeper rise at low $p_t^{\cut}$.  
Replacing the Lund string-fragmentation scheme 
by independent fragmentation and removing
resonance decays led to 
an  increase of the moments  for $p_t^{\cut}\simeq 0.5$ GeV,
but below this cut the moments approach constant values which are
larger than unity. An MC sample generated
without the colour-dipole parton 
emission and resonance decays, but  retaining  
the Lund string fragmentation, leads to an increase of the moments  
with decreasing $p_t^{\cut}$.

\begin{figure}
\begin{center}
\vspace{1.0cm}
\mbox{\psfig{file=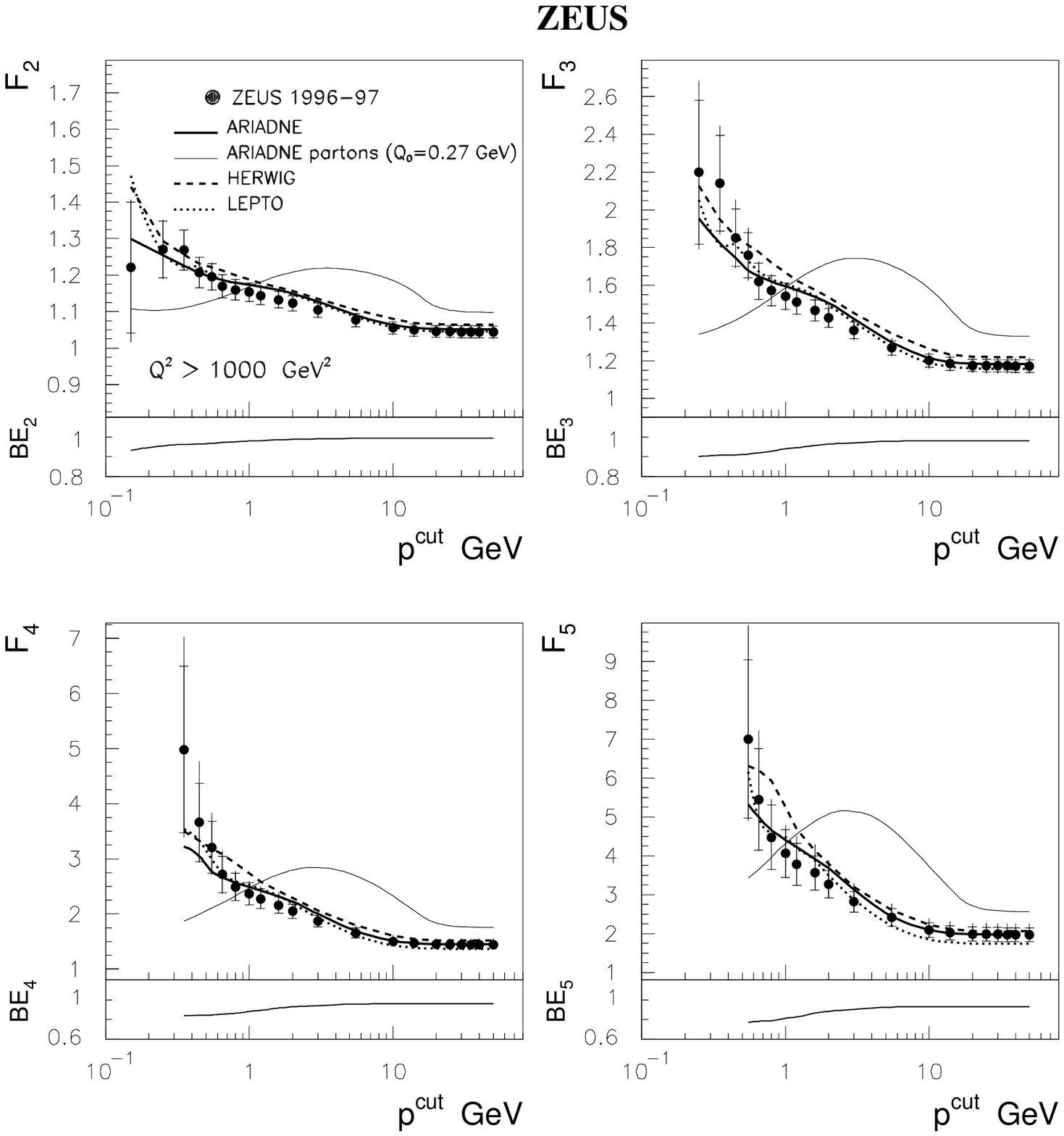, height=20.0cm}}
\vspace{-3.0cm}
\caption{
Factorial moments for charged particles
in the current region of the Breit frame  as a function of $p^{\cut}$,  
compared to Monte Carlo models.  The thin solid lines show the  parton-level
prediction of ARIADNE and the thick solid lines are the hadron-level  
predictions of the same model. 
The dashed and dotted lines show the hadron-level predictions
of HERWIG and LEPTO, respectively.   
The inner error bars
are statistical uncertainties; the outer are statistical and systematic
uncertainties  added in quadrature.
The correction factors for the BE  
effect are shown below each plot.
}
\label{fig:7}
\end{center}
\end{figure}

Figure~\ref{fig:7} shows   
the factorial moments for restricted $p$  values.
As was the case for the $p_t^{\cut}$ moments, there is 
no indication that the values approach unity at 
small  $p^{\cut}$ values. 
All MC models at the  hadron level agree with the data.
The BE interference has a large influence on this measurement, giving  
a $20\%$ correction.

The DLLA QCD calculations neglect  
energy-momentum conservation  in gluon
splittings, considering idealised jets at asymptotically  large energy. 
This effect  can
be taken into account in the Modified Leading Log Approximation; however, 
such  calculations have not yet been performed.
Therefore, it is important to compare the experimental results with
the parton shower of  MC models that explicitly include  
energy-momentum conservation and the complete parton-splitting functions. 
In addition, such comparisons   
allow the investigation of the contributions  from hadronisation. 

The moments restricted in $p_t$ and $p$ variables
were determined  from the ARIADNE parton-level,
the physics implementation of which strongly resembles the analytic
calculations \cite{ochs_pt}. The LPHD requires $Q_0$ to approach $\Lambda$.
The ARIADNE default value of $\Lambda$ is 0.22 GeV, while
the transverse momentum cut-off $Q_0$ is $0.6$ GeV. Therefore, 
to satisfy the LPHD requirement, the parton predictions were  obtained 
after switching off the string fragmentation and lowering the $Q_0$ value in ARIADNE to $Q_0=0.27$ GeV.  
This value was chosen in order to insure that
the average multiplicity of partons
in the current region of the Breit frame equalled that of
hadrons, which is important for normalisation purposes.
This method is somewhat different to that used in 
a similar study  of the ARIADNE parton cascade for $\E$
annihilation events \cite{ochs_pt}.
In contrast to that approach, the present modification
of the model retains the original  default value of $\Lambda$,
to preserve  
the kinematics of first-order QCD processes in the Breit frame.

The parton-level ARIADNE for $p_t^{\cut}$ and $p^{\cut}$ moments
is shown as thin solid lines in Figs.~\ref{fig:6} and
~\ref{fig:7}. 
Agreement between the parton-level predictions and the
data for the lowest $p_t^{\cut}$ values can be excluded
at greater than $99\%$  confidence level for $q=2,3$  and at greater 
than $95\%$ confidence level for $q=4,5$. 
The data for all $p^{\cut}$ moments 
disagrees with the parton-level predictions by
many standard deviations. 
The MC predictions for partons
have  the trends expected from the analytic DLLA results,
showing  a decrease of
the moments for decreasing $p_t^{\cut}$ and $p^{\cut}$.
This is contrary both to the 
measurements and to the  
hadron-level MC predictions,  which show a rise of the moments. 
Analogous  differences between the hadron and parton levels
of ARIADNE  have also been
observed in  $\E$ annihilation \cite{ochs_pt}.

The parton predictions in the full current region (large 
$p_t^{\cut}$ and $p^{\cut}$) show  larger values of the moments than
the data, despite the tuning of the parton    
to the hadron multiplicities. In the case of $e^+e^-$,
the MC-tuned parton level describes  the hadron-level
predictions  for large momentum cuts \cite{ochs_pt}. 
It is likely that the discrepancy between
partons and hadrons in DIS is due to additional
non-perturbative effects related to the proton remnant.

\subsection{Angular moments} 

\begin{figure}
\begin{center}
\mbox{\epsfig{file=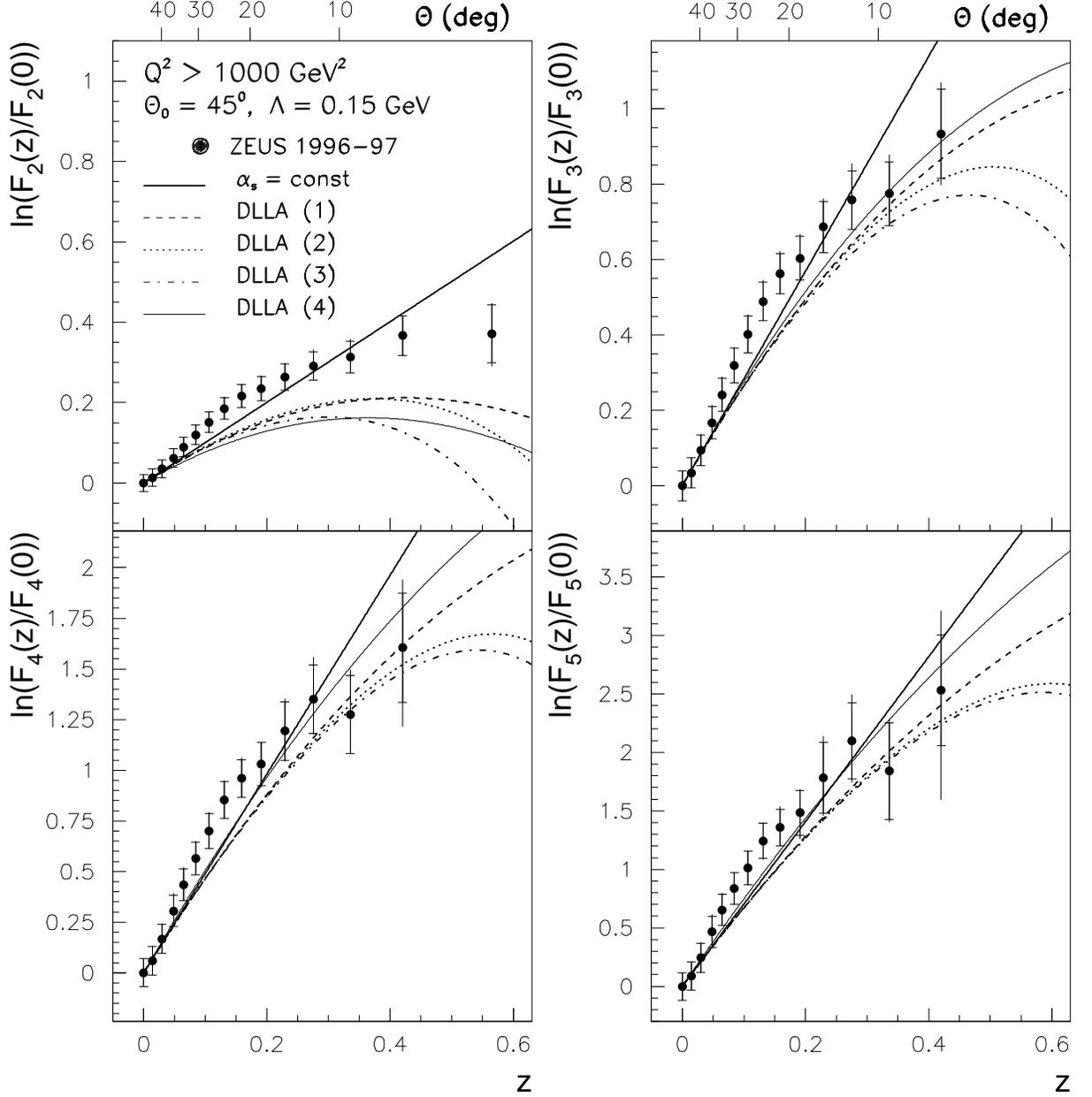, height=18.0cm}}
\caption{
$F_{q}(z)/F_{q}(0)$  $(q=2,\ldots ,5)$ as a function
of the scaling variable $z$ for $\Theta_0 = 45^o$
and  $\Lambda = 0.15$ GeV, compared to the 
QCD prediction of Eq.~(2) with different
parameterisations of $D_q$:     
$\alpha_s=const$ of Eq.~(4) (bold solid lines); 
(1) DLLA of Eq.~(5) (dashed lines);  (2)  
DLLA of Eq.~(6) (dotted lines); (3)  DLLA of Eq.~(7) 
(dash-dotted lines);  and (4)  DLLA with a correction
from the MLLA \cite{drem} (thin solid lines). 
The values of $\Theta$ are given at the top of the plots.   
The inner error bars
are statistical uncertainties; the outer are statistical and systematic
uncertainties  added in quadrature.
}
\label{fig:8}
\end{center}
\end{figure}

Figure~\ref{fig:8} shows the angular 
moments $F_q(z)/F_q(0)$. This represents the first measurement
of these quantities in DIS. The measurements are 
compared to DLLA QCD calculations 
in Eq.~(2),  after substitution of 
different forms of the R\'{e}nyi dimension
(see Eqs.~(4)-(7) ) and using the 
effective QCD parameter $\Lambda=0.15$ GeV as 
in LEP studies \cite{fluc_l3,fluc_del}.  
For DIS in the Breit frame, the energy $E=Q/2$ 
was chosen to define the $\as$ value,  
corresponding to the outgoing quark momentum in the quark-parton model. 
The factors used  to correct the data for 
the BE interference are close to unity for all angular moments (not shown).
A significant disagreement with the data is  observed, 
which becomes smaller for higher orders of the moments.
Although the error bars are clearly correlated,
even for the high-order moments,  
agreement between the predictions and the data can be excluded
for $z=$0.05-0.2 at greater than $95\%$ confidence level.
For running-$\as$ calculations, 
the best description is observed for high-order moments in the case of  
the DLLA result with  a correction from the MLLA.  
This result is similar to that obtained for 
$\E$ annihilation  \cite{fluc_l3, fluc_del}. It should be noted
that the present analysis,
however, uses a larger
angle $\Theta_0=45^o$, in order to increase the statistics, and 
that the average energy ($\la E\ra \simeq 23$ GeV) 
of the outgoing quark giving rise to the parton shower
is smaller than at LEP by a factor of two.

\begin{figure}
\begin{center}
\mbox{\epsfig{file=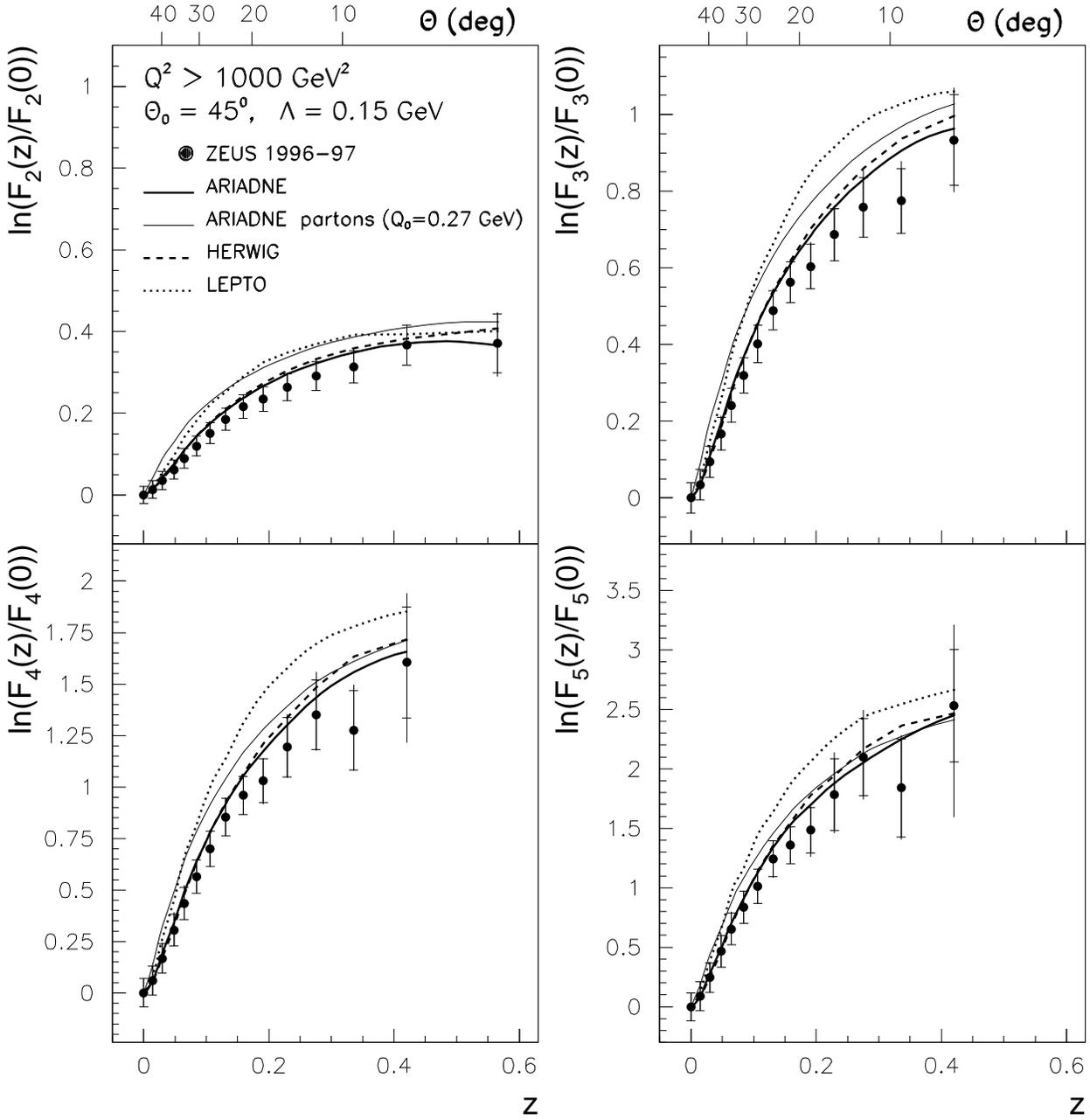, height=18.5cm}}
\caption{
The  data of  Fig.~\ref{fig:8} compared to different  
Monte Carlo models at the hadron level: ARIADNE (bold solid lines), 
HERWIG (bold dashed lines) and LEPTO (bold dotted lines).
The thin solid lines show the  parton-level
prediction of ARIADNE with $Q_0=0.27$ GeV. 
}
\label{fig:9}
\end{center}
\end{figure}

Figure~\ref{fig:9}  
shows  the comparisons of the angular moments
with Monte Carlo models at the hadron level.  
All models reproduce  the trends of the data but somewhat 
overestimate the strength of the correlations;  
however, LEPTO fails to give a reasonable
description of the angular moments.

The parton-level ARIADNE predictions
for the angular moments are shown in Fig.~\ref{fig:9}.
The ratio $F_q(z)/F_q(0)$ predicted by ARIADNE 
at the parton level (thin lines)
overestimates  the data, but is 
closer than the analytic DLLA calculations for the
second-order moment.
This comparison with the MC parton level and similar studies
of $\E$ annihilations  at LEP \cite{fluc_l3,fluc_del} 
imply that MLLA analytic calculations are necessary to
give a better description of the measured angular moments.

\section{Conclusions}

Multiplicity moments have been studied in deep inelastic scattering
at $Q^2>1000$ GeV$^2$
with the ZEUS detector at HERA.  
The moments measured in the current region of the Breit
frame show a strong rise as the size of the restricted intervals in
$p_t$, $p$ and $\Theta$  decreases. 
Monte Carlo models, 
which include hadronisation effects, reproduce the 
correlation pattern of the hadronic final state, 
although there are some discrepancies with the data.

The analytic QCD calculations for the angular moments
agree qualitatively with the data, but there are discrepancies
at the quantitative level. An inadequate  
treatment of energy-momentum conservation in gluon splittings,
which is important at low energies,
is likely  to be responsible for the discrepancies between  the
analytic QCD expectations and the data.
The parton level of ARIADNE, which does  
not suffer from the  limitations
of the DLLA calculations,  indicates  that 
the discrepancies can be reduced. MLLA calculations
are required to provide an accurate description of the angular moments.
Similar studies of the angular moments measured at LEP  
also show that the analytic perturbative
calculations need to be improved.

In contrast to the angular fluctuations,
the moments calculated in restricted $p_t$ and $p$ regions show
significant discrepancies, both for the DLLA calculations
and for the parton-level of ARIADNE.
This is most clearly seen for the $p_t^{\cut}$ moments;
while the analytic calculations
and ARIADNE predictions for partons decrease and approach unity for small
$p_t$ regions, the data show a steep rise.
This is the  first indication that perturbative QCD, 
in conjunction with the LPHD hypothesis, fails  on a qualitative level
to describe the hadronic  multiplicities. 
Monte Carlo studies of the hadronisation effects
indicate  that this is due to
violation of the LPHD hypothesis for many-particle densities
measured in the regions restricted in the transverse momenta
with respect to the direction of the outgoing quark.

\section*{Acknowledgements}
The strong support and encouragement of the DESY Directorate have
been invaluable, and we are much indebted to the HERA machine group
for their inventiveness and diligent efforts.  The design,
construction and installation of the ZEUS detector have been made
possible by the ingenuity and dedicated efforts of many people from
inside DESY and from the home institutes who are not listed as authors.
Their contributions are acknowledged with great appreciation.
We thank  W.~Ochs and  J.~Wosiek   
for helpful  discussions.

\newpage
\clearpage
\pagebreak
 
{}

\newpage 



\newpage


\newpage 

\newpage

\newpage

\newpage 

\end{document}